\documentclass[12pt,fleqn]{article}

\usepackage{epsfig}
\usepackage{amssymb}
\usepackage{amsthm}
\usepackage{amsmath}

\newtheorem{teo}{Theorem}
\newtheorem{lem}[teo]{Lemma}

\newcommand{\Eq}[1]{(\ref{#1})} 		    
\newcommand{\Th}[1]{Theorem~\ref{#1}} 		
\newcommand{\eqs}[2]{(\ref{#1},\ref{#2})} 
\newcommand{\Sec}[1]{\S\ref{#1}} 	
\newcommand{\Fig}[1]{Fig.~\ref{#1}} 		

\newcommand{\half} {\frac12}

\newcommand{\Epsfig}[5] {
					   \begin{figure}[#2]
						   \epsfxsize=#5
						   \centerline{\epsfbox{#1}}
						   \caption{\small { #3 }\label{#4}}
					   \end{figure}
}

\def \fix {\mbox{Fix}}
\def \dilog {\mbox{dilog}}

\title{Heteroclinic orbits and transport in a perturbed integrable Suris map\footnote{
	A version of this paper was originally written in 1996, but never
	published.  Subsequently, the methods of this paper have been used and
	extended, for example by Delshams and Ram\'\i rez-Ros
	\cite{Delshams4}.}
}
\author{
	H\'ector E. Lomel\'\i \\
	Department of Mathematics\\
	Instituto Tecnol\'ogico Aut\'onomo de M\'exico\\
	Mexico, DF 01000\\
	lomeli{@}gauss.rhon.itam.mx
\medskip \\
	James D. Meiss\thanks
	{JDM was supported in part by NSF grant number DMS-9971760.}\\
	Department of Applied Mathematics\\
	University of Colorado\\
	Boulder, CO 80309\\
	jdm{@}boulder.colorado.edu}

\begin{document}
\maketitle
\begin{abstract}
	Explicit formulae are given for the saddle connection of an integrable
	family of standard maps studied by Y. Suris.  When the map is perturbed
	this connection is destroyed, and we use a discrete version of
	Melnikov's method to give an explicit formula for the first order
	approximation of the area of the lobes of the resultant turnstile. 
	These results are compared with computations of the lobe area.
	
\center\medskip\noindent
	{\bf AMS classification scheme numbers:}
		34C35,34C37,58F05,70H15,70K99	

\center\medskip\noindent
	{\bf Keywords:}
		Integrable Maps, Melnikov Method, Transport, Twist maps
\end{abstract}

\section{Introduction} \label{sec:intro}

Standard maps are area-preserving diffeomorphisms of ${\mathbb T}\times
{\mathbb R}$ given by
\begin{equation} \label{stdMap}
    (\theta',r') = f(\theta,r) = (\theta+r+V'(\theta),r+V'(\theta)) \;,
\end{equation}
where the potential, $V$, is periodic, $V(\theta+1)=V(\theta)$.  In
this paper we study the standard map $f_{\delta}$ given by \Eq{stdMap}
with the potential\footnote
{
	The potential can also be written
	\[
		V(\theta) = {\frac{1}{\pi^2}}\Re\left[{\dilog(1+\delta) -
				\dilog(1+\delta {\rm e}^{2\pi i\theta })}\right] \;.
	\]
	where the dilogarithm is defined by
    $ \dilog(x)\equiv \int_1^x {\frac{\log(z)}{1-z}} dz$.
}
\begin{equation} \label{surispot}
    V_{\delta}(\theta ) = -{\frac 2 \pi}\int_0^\theta dt\ {\tan}^{-1}
           \left(
		     {{\frac{\delta \sin(2\pi t)}{1+\delta \cos\ (2\pi t)}}}
		   \right) \;.
\end{equation}
Suris \cite{Sur, Meiss1} showed that $f_{\delta}$ is integrable with
integral
\[
    I_\delta(\theta,r)=\cos\pi r+\delta\cos\pi(2\theta-r) \;, 
\]
i.e., $I_\delta\circ f_\delta=I_\delta$. Contours of $I_\delta$ are 
shown in \Fig{fig:invariant}.  The map is integrable for any $\delta$; 
however, we will consider the case $0<\delta<1$, as the topology of 
the saddle connections changes at $\delta=1$.  For $0<\delta<1$ the map 
$f_\delta$ has hyperbolic fixed points at $ z_a = (-\half,0)$ and $ 
z_b = (\half,0)$, that are connected by two saddle connections, 
forming the upper and lower separatrices of the fixed point resonance. 

\Epsfig{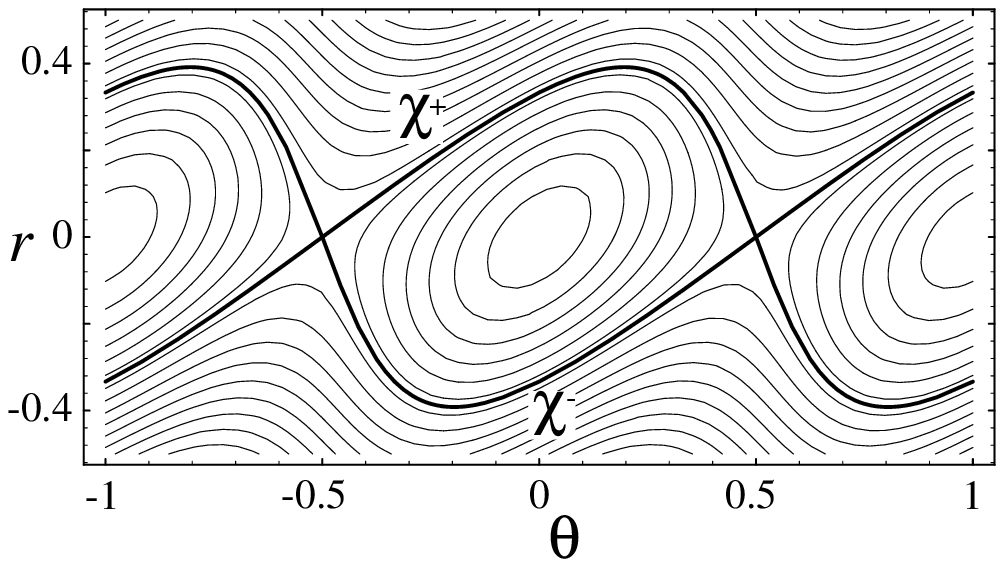}{ht}
   {Some contours of $I_\delta$, for $\delta=1/3$.}
   {fig:invariant}{9cm}

Standard maps are examples of monotone twist maps (for a review, see
\cite{Meiss1}) defined by the geometrical {\em twist condition} that vertical
lines tilt to the right upon iteration, $\frac{\partial
\theta'}{\partial r} >0$.  Such maps have Lagrangian generating
functions, $S(\theta, \theta')$, which generate the map implicitly
through the eq\-uation $dS(\theta,\theta') = r' d\theta' - r d\theta$,
or equivalently
\begin{equation} \label{genfun}
      r = - \partial_1 S(\theta,\theta') \;, \quad
      r'= \partial_2 S(\theta,\theta') \;,
\end{equation}
where the subscripts indicate derivatives with respect the first and
second arguments of $S$.  The twist condition implies that
$\partial_{1}\partial_{2}S < 0$.  We assume that our
twist map has {\it zero net flux}, which is equivalent to
$S(\theta+1,\theta'+1) = S(\theta,\theta')$.  For maps of the standard
form, \Eq{stdMap}, the generating function is
\begin{equation} \label{stdgen}
      S(\theta,\theta') = {\frac12} (\theta' - \theta)^{2} + V(\theta) \;.
\end{equation}
We will let $S_\delta$ denote the generating function for the Suris 
map $f_\delta$.

Since the Suris map is integrable and has twist, Birkhoff's theorem
\cite{Meiss1} implies that the saddle connection between the
equivalent points $z_a$ and $z_b$ is the graph of a function $r =
\chi(\theta)$, for $\theta \in {\bar{\cal U}}= [-\half, \half]$.  The
dynamics \Eq{stdMap} restricted to the saddle connection defines a circle
diffeomorphism $$\theta \mapsto h(\theta)=\theta+\chi(\theta) +
V'(\theta).$$

Our goal is to show that, when the Suris map is perturbed, the stable 
and unstable manifolds of the continuation of the fixed points 
intersect transversely.  We use a version of Melnikov's method 
\cite{Mel} for twist maps 
\cite{Easton84,Glasser,Delshams1,Lom1,Lom2,Lom0,Delshams3}.  The 
system is perturbed by adding to the original generating function any 
$C^2$ periodic function $P(\theta,\theta')$.  We will assume for 
simplicity that $P$ does not move the fixed points.  Then for small 
enough $\epsilon$, $S_\delta+\epsilon P$ is a twist generating 
function, and the corresponding map $f_{\delta,\epsilon}$ has 
hyperbolic fixed points at $z_a$ and $z_b$.  As was shown in 
\cite{Lom2}, if the Melnikov potential, given by
\begin{equation}\label{eq:melpot}
    L(\theta)=\sum_{t=-\infty}^{\infty} 
    P(h^t(\theta),h^{t+1}(\theta)) \;,
\end{equation}
has a nondegenerate critical point in ${\cal U}$, then the manifolds
$W^u(z_a,f_{\delta,\epsilon})$ and $W^s(z_b,f_{\delta,\epsilon})$
intersect transversely for $\epsilon$ small enough (the same
conclusion is valid for $W^s(z_a,f_{\delta,\epsilon})$ and
$W^u(z_b,f_{\delta,\epsilon})$).  It is known that the series for $L$
converges absolutely and uniformly to a $C^2$ function on compact
subsets of ${\cal U}$.  As an example below, we consider the case
$P(\theta,\theta')=\cos^2\pi\theta$.

Thus the potential $L$ provides sufficient conditions for the
transversal destruction of the saddle connection, just as in the
classic applications of the Melnikov integral.  In \Sec{sec:melnikov}
we formulate a slightly stronger version of the previous result
(\Th{thm:mel}).

The Melnikov potential also has a physical interpretation: the
difference between values of $L$ at neighboring critical points is
the area of the turnstile lobe that they delineate.  The area of the
lobe gives a coordinate independent measure of the separation of the
perturbed stable and unstable manifolds as well as the flux from one
region to another (cf. \cite{Meiss1}).

The sum for the order $\epsilon$ approximation for the lobe area is
calculated explicitly in \Sec{sec:surismap} and compared with numerical
results in \Sec{sec:numerical}.  We summarize our results, as
\begin{teo}[Main Theorem] \label{thm:main}
	Let $S_\delta$, given by \Eq{stdgen} with potential
	$V_{\delta}$ \Eq{surispot}, be the generating function for the
	integrable Suris map $f_\delta$.  Then
	\[
		S_{\delta,\epsilon}=S_\delta+\epsilon \cos^2\pi\theta
	\]
	generates a twist map $f_{\delta,\epsilon}$.  For all $0<\delta<1$ and
	$\epsilon$ small enough there are two neighboring, primary
	heteroclinic orbits from $z_a = (-\half,0)$ to $z_b = (\half,0)$.  The
	lobe defined by these two orbits has an area given by
	\begin{equation} \label{theoarea}
	   A(\delta,\epsilon) = \epsilon \Gamma(\nu) +O(\epsilon^2) 
	\end{equation}
	where $\nu \equiv (1-\sqrt{\delta})/(1+\sqrt{\delta})$ and
	\begin{equation}\label{gamma}
	 \Gamma(\nu)   \equiv 1+8\sum_{k=1}^\infty \frac{(-1)^kk\nu^k}{1+\nu^k}
	                = \vartheta_{4}^{4}(0,\nu) \;,
	 \end{equation}
	 where $\vartheta_{4}(z,q)$ is the Jacobi theta function.
\end{teo}

\noindent
Note that the multipliers of the hyperbolic fixed point of
$f_{\delta}$ are $\nu$ and $1/\nu$, so that $\nu$ is a natural
parameter to use (see e.g., \cite{Delshams3}).

The series $\Gamma(\nu)$ is rather intriguing.  It is an analytic
function of $\nu$ on the interval $\{0<\nu<1\}$, and approaches zero
rapidly as $\nu$ increases.  In fact, though \Eq{gamma} implies that
$\Gamma(\nu)$ is strictly positive, it approaches zero exponentially
fast:
\[
    \Gamma(\nu)\sim\left(\frac{4 \pi}{\log(1/\nu)}\right)^2
	\exp\left(\frac{-\pi^2}{\log(1/\nu)}\right)
\]
as $\nu\to1^-$, or equivalently $\delta\to 0^+$.  However, note that 
\Eq{theoarea} is only valid for fixed $\delta$ as $\epsilon \to 0$.  
If these parameters both approach zero, for example according to some 
relation $\delta = \delta(\epsilon)$, then the analysis becomes much 
more difficult.  In particular when $O(\epsilon)$ term in 
\Eq{theoarea} is exponentially small in $\delta$, it could easily be 
dominated by terms that are formally higher order in $\epsilon$.

\section{ Melnikov potential for twist maps }\label{sec:melnikov}
 
In this section we review the derivation of the Melnikov potential for
twist maps \cite{Lom2}.  We begin with a $C^2$ Lagrangian generating
function $ \tilde S(\theta,\theta')$ that satisfies the twist condition and
has zero net flux.  It gives a map of the annulus implicitly through
\Eq{genfun}.  Alternatively, let the action $W$ of a
sequence $[\theta] = \{ \theta^{i},\theta^{i+1},\ldots,\theta^{j}
\}$, be
\[
     W [\theta] = \sum_{t=i}^{j}{ \tilde S(\theta^{t},\theta^{t+1})} \;.
\]
Then, an orbit of the map that begins at $\theta^{i}$ and ends at
$\theta^{j}$ corresponds to a critical point of $W$ under variation
with respect to the interior points \cite{Meiss1}.  The corresponding
momenta are then defined through \Eq{genfun} as $r^{t} = - \partial_1
 \tilde S(\theta^{t},\theta^{t+1})$.  Thus, for example, a point
$(\theta_a,r_{a})$ is a fixed point of the map if and only if
$\theta_a$ is a critical point of $ \tilde S(\theta,\theta)$, and $r_{a}$ is
defined through \Eq{genfun}.

The distance between the perturbed stable and unstable manifolds of
any twist map with a saddle connection can be computed using the
variational principle \cite{MM88,Delshams2}.  We begin with a map
$ \tilde f $ generated by $ \tilde S$.  Suppose that $ \tilde f$ has two hyperbolic
fixed points $z_{a}=(\theta_a,r_a)$ and $z_{b}=(\theta_b,r_b)$, and
there is a saddle connection defined by the graph of a function
$\chi(\theta)$ on the interval ${\cal U} = (\theta_{a},\theta_{b})$
between these points.  A diffeomorphism $h:{\bar{\cal U}}\to
{\bar{\cal U}}$ is induced by the restriction of the map to the saddle
connection:
\begin{equation}\label{eq:hdef}
       \tilde f(\theta,\chi(\theta))=(h(\theta),\chi(h(\theta))) \;.
\end{equation}

Let $P$ be a $C^{2}$ function with zero net flux.  Then the function
\[
      \tilde S_\epsilon(\theta,\theta')= \tilde S(\theta,\theta')+
                                \epsilon \tilde  P(\theta,\theta')
\]
generates a twist map $f_\epsilon$ for small enough $\epsilon$.  
Since hyperbolic points are nondegenerate critical points of the 
action \cite{MM83}, the perturbed map will have nearby hyperbolic fixed 
points for small enough $\epsilon$.  A simple case occurs when $\theta_a$ is a 
critical point of $ \tilde P(\theta,\theta)$ as well as of 
$ \tilde S(\theta,\theta)$ since it is then a critical point of 
$ \tilde S_{\epsilon}(\theta,\theta)$ as well.  Thus the fixed points of 
$ \tilde S_{\epsilon}$ will have unchanged configurations, but their momenta will 
be modified according to \Eq{genfun}.

The action gives useful formulae for the areas of regions for twist
maps \cite{MMP2}.  We will use one such relation to obtain the
Melnikov potential: a relation between the graph $\chi$ and the action
of orbits on the stable and unstable manifolds of a hyperbolic fixed
point \cite{Tab}.  Let $z^{0}$ be a point on the unstable manifold of
a fixed point $z_a$ that is close enough to $z_{a}$ so that the
segment of $W^{u}(z_{a})$ containing $z^{0}$ is given by a graph
$(\theta,\chi(\theta))$.  Let $(\theta^{t}, \chi(\theta^{t})), t \le
0$, be the preorbit of this $z^{0}$.  Defining the backward action
difference as
\[
	\Delta W^{B}(\theta^{0}) = \sum_{t=-\infty}^{-1} 
	        \left[
			      \tilde S(\theta^{t},\theta^{t+1})-  \tilde S(\theta_a,\theta_a)
		     \right] \;,
\]
then the unstable manifold is defined by the graph of the function
\[
    \chi^{u}(\theta) = \frac{\partial}{\partial \theta} \Delta W^{B}\;.
\]
A corresponding formula for the forward action of an orbit on the 
stable manifold yields a formula for the graph of an initial segment of 
the stable manifold, $\chi^{s}$, of $z_b$:
\begin{eqnarray}
    \Delta W^{F}(\theta^{0})   =&  - \sum_{t=0}^{\infty} 
	             \left[
		        \tilde S(\theta^{t},\theta^{t+1})-  \tilde S(\theta_b,\theta_b)
		     \right]
	             \;, \nonumber \\
    \chi^{s}(\theta) =& \frac{\partial }{\partial \theta} \Delta W^{F} \;.
\end{eqnarray}

The difference between these two actions leads to the
Melnikov-like formula for the transversal intersection of these 
manifolds. To summarize:

\begin{teo} \label{thm:mel}
	Let $ \tilde S$ be the generating function for a twist map $ \tilde f$ that has
	two hyperbolic fixed points $z_{a}$ and $z_{b}$ with a saddle
	connection given by $r=\chi(\theta)$ for $\theta \in {\cal U} =
	(\theta_{a},\theta_{b})$.  Let $ \tilde f$ induce a diffeomorphism
	$h(\theta)$ on the connection.
	  Let $ \tilde S_\epsilon =  \tilde S+\epsilon  \tilde P$
	generate the twist map $ \tilde f_\epsilon $, such that the perturbation $ \tilde P$
	has the following properties
	\begin{description}
		\item[a)] $ \tilde P(\theta_a,\theta_a)= \tilde P(\theta_b,\theta_b)=0$ 
		\item[b)] 
$\left.\frac{d}{d\theta}\right|_{\theta=\theta_a} \tilde P(\theta,\theta)=
		\left.\frac{d}{d\theta}\right|_{\theta=\theta_b} \tilde P(\theta,\theta)=0$
	\end{description}
	Then for $\epsilon > 0$ small enough:
	\begin{description}
		\item[i)] the perturbed map has two hyperbolic fixed points near 
$z_a$
	   	    and $z_b$; 
	    \item[ii)] the Melnikov potential
		\begin{equation}\label{laele}
				  L(\theta)=\sum_{t=-\infty}^{\infty} 
 \tilde P(h^t(\theta),h^{t+1}(\theta))
		\end{equation}
				 converges absolutely and uniformly to a $C^{2}$ 
function on 
				 ${\cal U}$; and
		\item[iii)] if $L$ has a nondegenerate critical point on ${\cal U}$,
					then the unstable and stable manifolds of the two 
fixed 
					points intersect transversely.
	\end{description}
\end{teo}

We can relate the Melnikov potential to the area of a lobe in the
stable and unstable manifolds using the action formula of \cite{MMP}. 
Suppose $p$ and $q$ are two neighboring heteroclinic orbits, for
example as shown in \Fig{fig:resonance}.  Then the area lobe is
determined by the difference in action between the orbits of $p$ and
of $q$ \cite{MMP}
\begin{equation} \label{deltaW}
    A =  \sum\limits_{ t=-\infty }^{\infty } 
         \tilde  S({\theta }_{q}^{t},{\theta }_{q}^{t+1})
       -  \tilde S({\theta }_{p}^{t},{\theta }_{p}^{t+1}) \;.
\end{equation}
The area $A$ is the signed area below the segment of $W^{s}$ between
$q$ and $p$ and above that of $W^{u}$.

\Epsfig{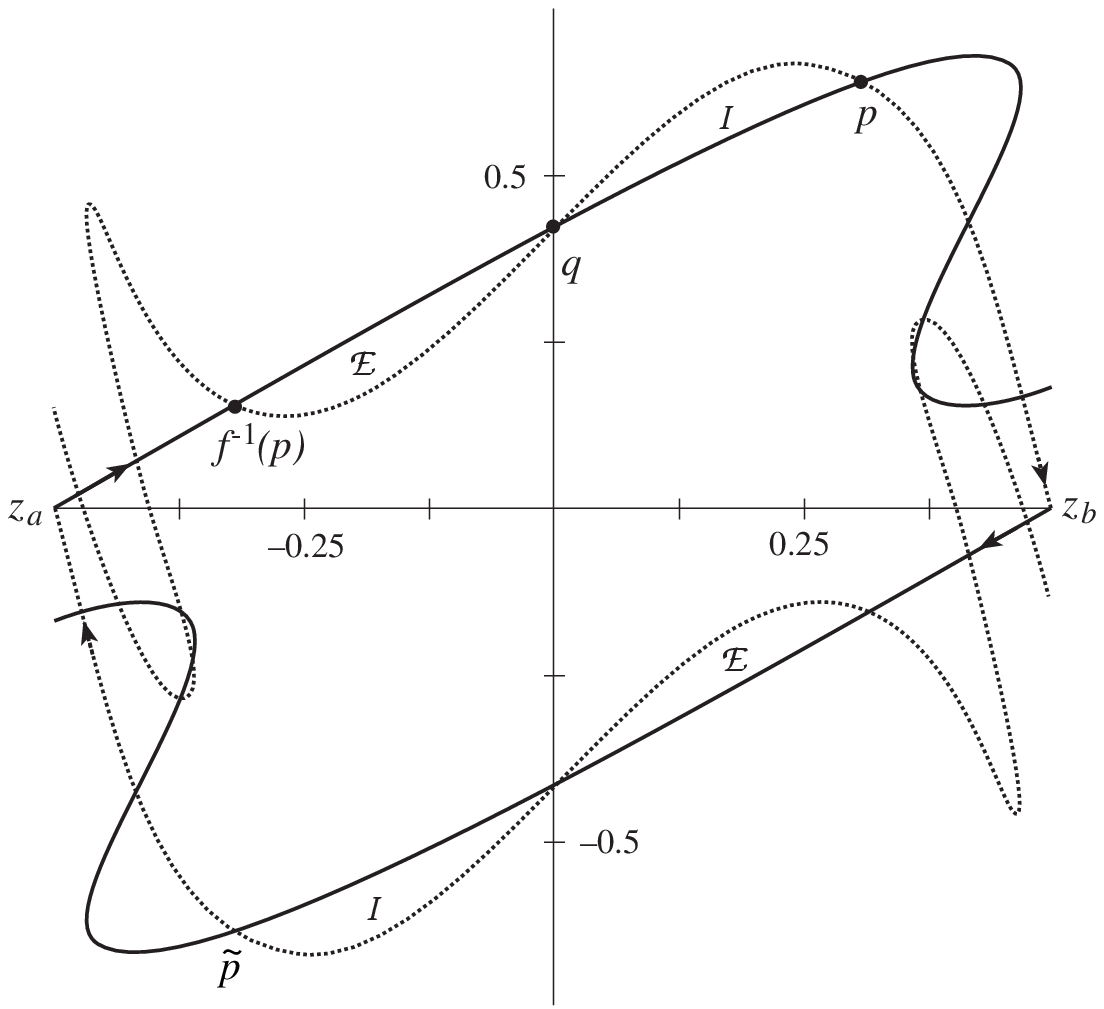}{ht}
     {Resonance for the perturbed Suris map with $\delta=0.5$
     and $\epsilon=0.05$. A pair of principal homoclinic points are 
     labeled $p$ and $q$, and the exit and incoming sets 
     are labeled $\cal{E}$ and $\cal{I}$.}
   {fig:resonance}{9 cm}

Since the action is stationary on an orbit and since the actions of
$p$ and $q$ are equal when there is a saddle connection, it is easy to
see that the area of the lobe is determined, to lowest order, by the
difference between two critical values of the Melnikov potential $L$
(cf. \cite{MM88, Delshams2}).  To summarize:
 
\begin{teo}
\label{thm:cab}
Let $ \tilde S_\epsilon$, $ \tilde f_\epsilon$, ${\cal U}$, $h$ and 
$L$ be defined as in \Th{thm:mel}.  Assume that $\theta_p$ and 
$\theta_q$ are two points in ${\cal U}$ such that
\begin{description}
\item[a)]$L'(\theta_q)=L'(\theta_p)=0$ ,
\item[b)]$L''(\theta_p)>0$, $L''(\theta_q)<0$,
\item[c)]$ L'(\theta) \neq 0$ for $\theta_q<\theta<\theta_p$.
\end{description}
Then the heteroclinic points $p$ and $q$ continue to heteroclinic
points of $f_{\epsilon}$, and the stable and unstable manifolds of
$f_\epsilon$ enclose a lobe with area
\[
	A(\epsilon)=\epsilon(L(\theta_q)-L(\theta_p))+O(\epsilon^2) \;.
\]
\end{teo}

\section{The Suris Map}\label{sec:surismap}

We will use the results of the previous section with 
$\tilde f=f_\delta$, $\tilde S=S_\delta$, 
$\tilde S_{\epsilon}=S_{\delta,\epsilon}$, $\tilde 
f_\epsilon=f_{\delta,\epsilon}$ and the particular perturbation 
$\tilde P(\theta,\theta')=\cos^2\pi\theta$.  We give formulas for the lobe area 
of the perturbed Suris map.  To find the saddle connection, we must 
find the diffeomorphism $h$ induced by the dynamics of the Suris map 
$f_{\delta}$ on the connection (\ref{eq:hdef}).  We will consider $h$ 
defined only on the set ${\cal U}=\{-\frac12<\theta<\frac12\}$.

We first give some properties of $h$.
\begin{lem}\label{funcion}
	Let $h$ be the diffeomorphism of ${\bar{\cal U}}$ given by
	\begin{equation}\label{difeofeo}
		 h_\nu(\theta)=\frac2\pi\arctan
		   \left(
			  \frac{(\nu+1)\tan(\frac\pi2\theta)+(\nu-1)}
			  {(\nu-1)\tan(\frac\pi2\theta)+(\nu+1)}
		   \right) \;\;,
	\end{equation}
	where $\nu=(1-\sqrt{\delta})/(1+\sqrt{\delta})$ and $0<\delta<1$.  Then 
$h_\nu$ satisfies
	\begin{description}
		\item[a)]$h^t_\nu=h_{\nu^t}$, for all $t\in\mathbb{Z}$.
		\item[b)]$h_\nu(-\theta)=-h_{\nu^{-1}}(\theta)$.
		\item[c)] $-\frac12$ is an stable fixed point and $\frac12$ is a 
		unstable fixed point of $h_{\nu}$.
		\item[d)] 
		$V'_\delta(\theta)=h_\nu(\theta)-2\theta+h_\nu^{-1}(\theta)$,
		for $\theta\in {\bar{\cal U}}$, for $V_{\delta}$ given by 
\Eq{surispot}.
	\end{description}
\end{lem}

\proof
A direct computation proves a) b) and c).  This stability properties
are implied by $h_{\nu}'(- \frac12) = \nu$ and $h_{\nu}'(\frac12) =
\frac{1}{\nu}$.  Since $\nu \in (0,1)$ this implies that the former is
stable and the latter is unstable.  For d), we will first show that
when $\theta\in {\bar{\cal U}}$,
\begin{equation}\label{lastar}
     \theta-h^{\pm 1}_\nu(\theta)=\frac2\pi\arctan
       \left(
         \frac{\pm\sqrt{\delta}\cos\pi\theta}{1 \mp \sqrt{\delta}\sin\pi\theta} 
       \right) \;\;,
\end{equation}
To see this, notice on one hand that
if $\theta\in {\cal U}$, then $h(\theta)\in {\cal U}$. Therefore
$-1 < \theta-h_\nu(\theta)  < 1$. On the other hand,
\begin{equation}\label{squares}
     \tan\left(\frac\pi2[\theta-h_\nu(\theta)]\right)=
       \frac{\sqrt{\delta}\cos\pi\theta}{1-\sqrt{\delta}\sin\pi\theta} \;\;.
\end{equation}
This implies \Eq{lastar} for the upper signs.  The substitution
$\theta\mapsto -\theta$ in equation \Eq{squares}, gives the lower signs. 
It is easy to see that if $\theta\in{\cal U}$, then \Eq{squares} is
positive and therefore $0<\theta-h_\nu(\theta)<1$.  We conclude that
for all $\theta\in {\cal U}$,
$-1<2\theta-h_{\nu^{-1}}(\theta)-h_\nu(\theta)<1$.  To finish the
proof we take tangent of the second difference and use the sum formula 
for tangent to obtain
\begin{eqnarray*}
 \tan\left({\frac\pi2}[2\theta-h_{\nu^{-1}}(\theta)-h_\nu(\theta)]\right) 
   &=& \frac{ 2 \delta\cos\pi\theta \; \sin\pi\theta}{1-\delta\sin^2\pi\theta+
       \delta\cos^2\pi\theta}\\
   &=& \frac{\delta\sin(2\pi\theta)}{1+\delta\cos(2\pi\theta)} \;.
\end{eqnarray*}
Since this last expression is 
$ - \tan(\frac{\pi}{2}V'_\delta(\theta))$, this completes the proof.
\qed

\medskip

With the help of lemma \ref{funcion}, we can give a description of the
intersection of the saddle connection between $(-\frac12,0)$ and
$(\frac12,0)$.  This connection is given by the functions
$\chi^{\pm}$, and is shown as the heavy curves in \Fig{fig:invariant}.
\begin{lem}
	Let $f_\delta$ be the twist map generated by $S_\delta$.
	Then $(-\frac12,0)$ and
	$(\frac12,0)$ are hyperbolic fixed points for $f_\delta$ and
	there are two saddle connections between them given by the graphs
		$r = \chi^{+}(\theta) = \theta - h_{\nu}(\theta)$, and
		$r = \chi^{-}(\theta) = \theta - h_{\nu^{-1}}(\theta)$. 

\end{lem}

\proof
Using Lemma \ref{funcion}, the map can be written in the form
\[
f_{\delta}(\theta,r)=(r+h_\nu(\theta)-\theta+h_{\nu^{-1}}(\theta),
 r+h_\nu(\theta)-2\theta+h_{\nu^{-1}}(\theta)) \;.
 \]
Therefore, if $r=\chi^-(\theta)=\theta-h_{\nu^{-1}}(\theta)$ then 
\[
    f_{\delta}(\theta,\chi^-(\theta))=(h_\nu(\theta),h_\nu(\theta)-
    \theta)=(h_\nu(\theta),\chi^-(h_\nu(\theta)) \;.
\]
Now since $\theta = -\frac12$ is an stable fixed point for 
$h_{\nu}$, this graph gives  the left going saddle connection.
In the same way setting $r=\chi^+(\theta)=\theta-h_\nu(\theta)$ gives 
\[
   f_{\delta}(\theta,\chi^+(\theta))=
     (h_{\nu^{-1}}(\theta),h_{\nu^{-1}}(\theta)-\theta)=
     (h_{\nu^{-1}}(\theta),\chi^+(h_{\nu^{-1}}(\theta)) \;.
\]
Since the map conjugates to $h_{\nu^{-1}}$ on this graph, this is 
clearly the right going saddle connection.
 \qed


\subsection{Proof of the Main Theorem}
We now can sketch the proof of the main theorem.  The analysis 
of the infinite series for the Melnikov potential relies on some 
summation formulae for elliptic integrals.

\proof 
Let $P(\theta,\theta')=\cos^2\pi\theta$.  It is clear that $P$
satisfies the conditions of \Th{thm:mel}.  Let $L$ be the Melnikov
potential, \Eq{eq:melpot}.  According to \Th{thm:mel} a sufficient
condition for transversal intersection of the perturbed manifolds is
that $L$ has a nondegenerate critical point on the interval
${\cal{U}}=\{-\half<\theta<\half\}$.  The graph of $L(\theta)$ over $\cal{U}$
is shown in \Fig{figure:L}.

\Epsfig{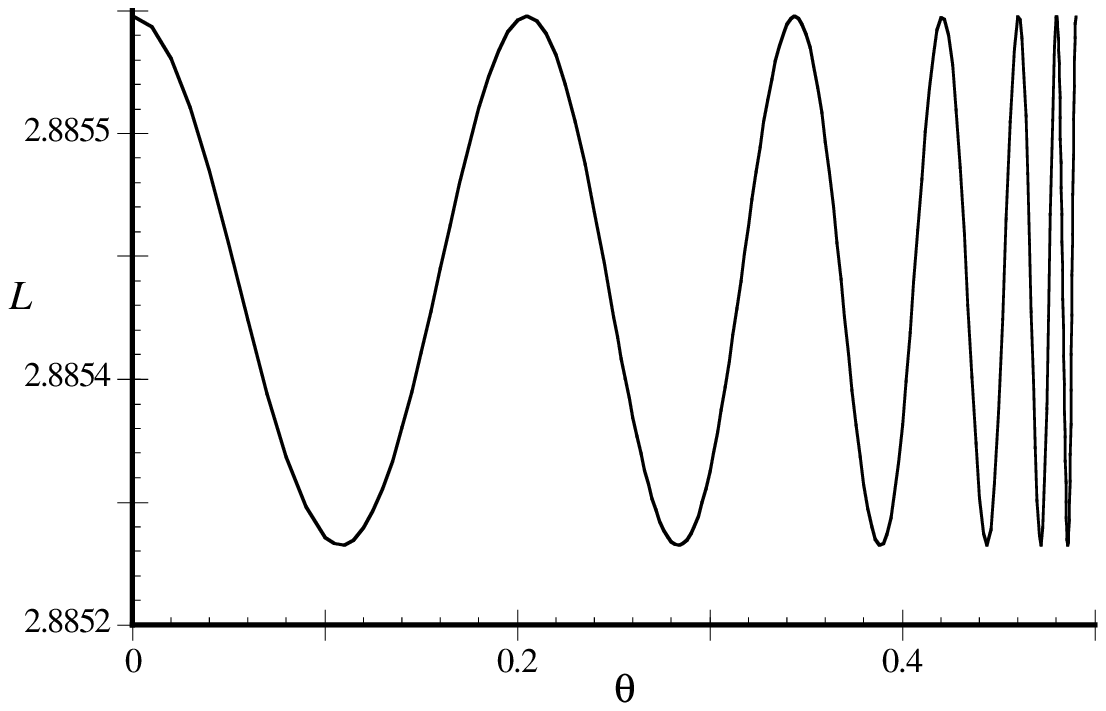}{ht}
       {Melnikov potential, $L$ for $\delta=1/2$.}
       {figure:L}{11 cm}

To proceed, we define $\bar L(z) = L(\frac{2}{\pi}\arctan(z))$. A 
little algebra gives $\bar L(z)=\sum_{t=-\infty}^\infty\alpha_t(z)$ 
where
\[
       \alpha_t(z)=
      {\frac{4 \nu^{2t}(1-z^2)^2}{((1-z)^{2} + \nu^{2t}(1+z)^{2})^2}} \;.
\]
Since $z = \tan(\frac{\pi}{2}\theta)$ is monotone increasing, we need only
show that there is a nondegenerate critical point of $\bar L$.  We are
going to show that $z_{q} = 0$ is a local maximum of $\bar L$ and
$z_{p} = (1-\sqrt{\nu})/(1+\sqrt{\nu})$ is a local minimum.

First, we write $\bar L$ as
\begin{equation}\label{delimon}
       \bar L(z)=\alpha_0(z)+\sum_{t=1}^\infty\{\alpha_t(z)+\alpha_{-
t}(z)\}\;\;.
\end{equation}
Now, since $\alpha_{t}(-z)=\alpha_{-t}(z)$, this implies that $\bar 
L$ is even, and so $\bar L'(0)=0$. Differentiation gives
\[
      \bar L''(0)=-8+\sum_{t=1}^\infty
         {\frac{128\left( 1 - 4{\nu^{2t}} + {\nu^{4t}}\right)
           {\nu^{2t}}}{{{\left( 1 + {\nu^{2t}} \right) }^4}}}
		    \equiv -8 \Gamma_0(\nu^2) \;.
\]
This expression defines a function $\Gamma_{0}$, which can be
rewritten by expanding the denominator and rearranging the sums as
\[
     \Gamma_0(q)=
       1+16\sum_{k=0}^\infty \frac{(-1)^kk^3q^k}{1-q^k} \;\;.
\]
This sum can be obtained from Fourier expansions of Jacobi elliptic 
functions which are written in terms of the nome, $q(k) 
=\exp\left(-\pi\frac{K'}{K}\right)$, where $K(k)$ is the complete 
elliptic integral, $K' = K(k')$ and ${k'}^{2}=1-k^{2}$ \cite{Byrd}.  
Recall that $q$ is a monotone increasing diffeomorphism of $k$ on 
$(0,1)$.  A Fourier expansion for elliptic functions 
\cite[formula 912.01]{Byrd} can be differentiated to show that
\[
     \Gamma_{0}(q) = \left(\frac{2}{\pi} k' K \right)^4  = 
     \vartheta_{4}^{8}(0,q) \;,
\]
where $\vartheta_{4}(z,q)$ is a Jacobi theta function \cite[formula 
1051.01]{Byrd}.  Therefore $\bar L''(0)<0$, for all $0<\nu<1$.  This 
shows that $\theta_q=0$ is a nondegenerate local maximum of $L$ for 
all $0<\nu<1$.

We now wish to show that $z_{p}$ is the local minimum. Rewrite $\bar L$ as
\begin{equation}\label{nieve}
       \bar L(z)=\sum_{t=0}^\infty\{\alpha_{t+1}(z)+\alpha_{-t}(z)\} \;\;.
\end{equation}
Since $\alpha_{t+1}(z_{p}+\zeta) = \alpha_{-t}(z_{p}-\zeta)$, $\bar L$
is even around $z_{p}$, and so $\bar L'(z_{p})=0$.  After some
algebra, the second derivative is
\[
      \bar L''(z_{p})=\sum_{t=0}^\infty
      {\frac{\left( 1 - 4{\nu^{1 + 2t}} + 
      {\nu^{2 + 4t}} \right){\nu^{1 + 2t}} } 
      {{{\left( 1 + {\nu^{1 + 2t}} \right) }^4}}} \;\;.
\]
Notice that $\bar L''(z_{p})= \Gamma_0(\nu^2)-\Gamma_0(\nu)$.
Once again, using Fourier series for elliptic functions, we find
\[
     \Gamma_0(q^{2}) = {k'}^{2} \left(\frac{2}{\pi} K\right)^4 \;,
\]
so that
\begin{eqnarray*}
       \bar L''(z_{p}) &=& \left(\frac{2}{\pi} K\right)^4 (k{k'})^{2} \;,
\end{eqnarray*}
and therefore ${\bar L}''(z_{p})>0$, for all $0<\nu<1$.  This shows 
that $\theta_p$ is a nondegenerate local minimum of $L$, for all $0<\nu<1$.
  
Using \Th{thm:cab}, we conclude that a lobe of area
\[
      A(\delta,\epsilon) = \epsilon\left(L(\theta_q)-L(\theta_p)\right)
                           +O(\epsilon^2)
\]
is enclosed by the stable and unstable manifolds, where $\theta_p$ and 
$\theta_q$ are given above.  Finally, we use \eqs{delimon}{nieve} to 
obtain
\begin{eqnarray*}
     L(\theta_q)={\bar L}(0)
           &=& 1+8\sum_{t=1}^\infty \frac{\nu^{2t}}{(1+\nu^{2t})^2} \;,\\
     L(\theta_p)={\bar L}(z_{p})
           &=& 8\sum_{t=0}^\infty \frac{\nu^{2t+1}}{(1+\nu^{2t+1})^2} \;.
\end{eqnarray*}
Subtracting these two, and expanding the series we obtain
\[
   L(\theta_q)-L(\theta_p) = 1+8\sum_{k=1}^\infty \frac{(-
1)^kk\nu^k}{1+\nu^k}=\Gamma(\nu)\;\;,
\]
where $\Gamma(\nu)$ was defined in (\ref{gamma}). Finally, this 
result can be explicitly written as
\begin{equation}\label{gamma2}
     \Gamma(q) = \left(\frac{2}{\pi} K {k'} \right)^{2} 
               = \vartheta_{4}^{4}(0,q) \;,
\end{equation}
upon differentiating the Fourier series \cite[908.51]{Byrd} once.
We have therefore obtained \Eq{theoarea}. \qed
 
\medskip

The explicit formula for the first order approximation of the area is 
compared with numerical computations in Sec.  \ref{sec:numerical}.

\subsection{Exponentially small behavior}\label{sec:small}

In this subsection we investigate the asymptotics of the sum,
\Eq{gamma}, as $\delta \to 0$.  Our result is summarized as:

\begin{lem}\label{nini}
	Let $\Gamma(\nu)$ be defined by \Eq{gamma}, then
	\begin{equation}\label{expsmall}
	   \Gamma(\nu) \sim \left( \frac{ 4 \pi}{\log(1/\nu)}
						 \right)^2 
							  \exp \left(\frac{-
\pi^2}{\log(1/\nu)} \right)
		\mbox{  as  } \nu \to 1^- \;.
	\end{equation}
\end{lem}
\proof We use the expression \Eq{gamma2} in terms of the elliptic 
integral $K$ and the nome $q$.  Recall that $q \to 1^{-}$ as $k \to 
1^{-}$ and $k' \to 0^{+}$.  The needed asymptotic forms are 
\cite[formulae 112.01 and 112.04]{Byrd}
\begin{eqnarray*}
    K(k) &\sim&  \log(\frac{4}{k'})  \;,\\
    k'   &\sim&  4 \exp \left(- \frac{\pi^{2}}{2 \log(1/q)} \right ) \;,
\end{eqnarray*}
as $ k \to 1^-$.
Putting these into \Eq{gamma2} gives
the promised result.
\qed

We again note that \Eq{expsmall} does not imply that the area itself
is necessarily exponentially small: \Eq{theoarea} is only valid for
fixed $\delta$ as $\epsilon \to 0$.  Nevertheless we will see in
\S\ref{sec:numerical} that \Eq{expsmall} agrees remarkably well with
our numerical calculations of the lobe area.

\subsection{Anti-Integrable Limit} \label{sec:anti}

It is much easier to obtain an expression for the lobe area for large 
large $\epsilon$.  Our main purpose is to have an expression that is 
valid in the regime of large $\epsilon$ where numerical calculations 
are difficult.  This expression is easy to obtain using the 
``anti-integrable limit" \cite{MM91}.  For simplicity, in this section 
we assume that $P$ depends only upon $\theta$ in this section.  In 
this case the anti-integrable limit is obtained by scaling the action 
by $\epsilon^{-1}$, to get
\[
	\hat{S} = S(\theta,\theta^\prime)/\epsilon = P(\theta) + \epsilon^{-1} 
S_\delta \;,
\]
and then setting $\epsilon^{-1}=0$.
The point is that when $\epsilon \gg 1 > \delta$ the points on the two 
heteroclinic orbits 
are all found in a neighborhood of the critical points of the potential 
$P = \cos^2(\pi \theta)$, i.e., at $\theta = m/2$. 
In the anti-integrable limit, an orbit consists solely of a sequence of 
configuration points sitting at these critical points. 
Thus the two heteroclinic orbits are given by the sequences 
\[
       \{\theta_p^t\} =  \{...,-\half,-\half,\half,\half,...\} \;\;, \;\;\;
       \{\theta_q^t\} =  \{...,-\half,-\half,0,\half,\half,...\}
\]
The action is stationary on an orbit, thus
to first order in the small parameter $\epsilon^{-1}$,
the change in the orbit with $\epsilon$ can be ignored in the 
action difference to give
\[
    \frac{1}{\epsilon} A = \sum\limits_{t=-\infty}^{\infty} 
                            \left( 
                               \hat{S}(\theta_q^t,\theta_q^{t+1})  - 
                               \hat{S}(\theta_p^t,\theta_p^{t+1}) 
                             \right) + O(\epsilon^{-2})\;.
\]
For the Suris map, this yields
\begin{equation} \label{antiI}
     A(\delta,\epsilon) = \epsilon - \frac{1}{4} - \frac{1}{\pi^{2}}
         \left[ 
             \dilog(1+\delta) - \dilog(1-\delta)  
         \right] + O(\epsilon^{-1})
\end{equation}
We compare this result with the numerical calculations in 
\Sec{sec:numerical}.

\section{Numerical comparison} \label{sec:numerical}
	
In this section, we compare the theoretical results with numerical 
computations of the lobe area.  The task is to find the actions of the 
two homoclinic points $p$ and $q$.  For this task we use the symmetry 
of the Suris map.  The action difference between the orbits of $q$ and 
$p$ gives the lobe area, \Eq{deltaW}.

The Suris map is reversible: it is conjugate to its inverse by an 
involution $R$: $R^2 = I$ and $Rf _\delta R = f^{-1}_\delta $,
 where the reversor is
\begin{equation} \label{reversor}
         R:\ (\theta ,r)\ \to \ (-\theta ,r\ +V'(\theta )) \;.
\end{equation}
Fixed sets, $\fix(R) = \{(\theta,r): \theta =0\}$ and $\fix(f_\delta R) =
\{(\theta,r) : r = 2\theta\}$ of the reversor are important because
points $q \in W^u(z_a) \cap \fix(R)$ or $q \in W^u(z_a) \cap \fix(f_\delta R)$
are heteroclinic from $z_{a}$ to $z_{b}$ \cite{SDM}.  Thus to find a
heteroclinic orbit it is sufficient to do a one dimensional search for
points on the unstable manifold that intersect one of these fixed
sets. 

\Epsfig{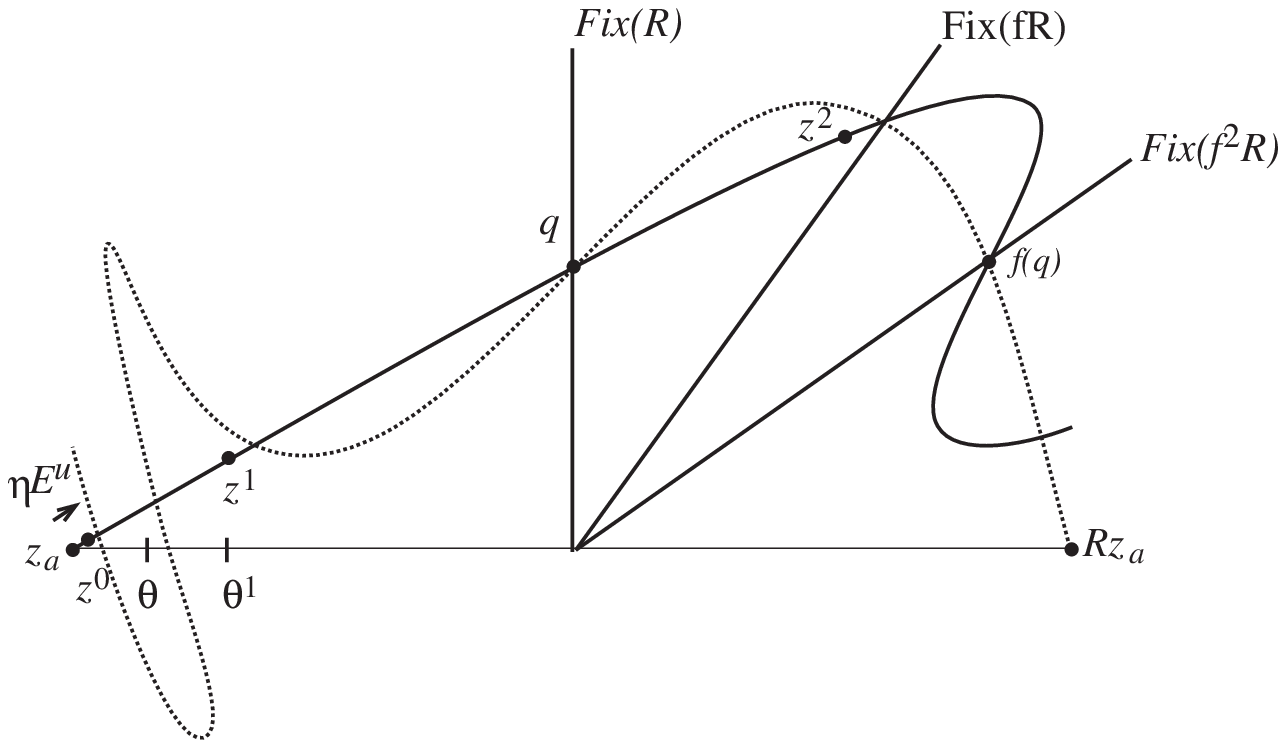}{ht}
       {Finding some heteroclinic orbits using symmetry. Shown are the 
		fixed sets $\fix(f^{t}R)$, for  $t = 0,1,2$
		and a rough example of initial guess 
		$z_{0}$, leading to $t_{c}=1$. }
        {fig:symmetry}{9cm}
	
To find these orbits numerically we move along $W^u(z_a)$ to the first
points that intersect $\fix(R)$ and $\fix(fR)$, respectively.  For
example, to find $q \in \fix{R}$, let $z^0 = (\theta^0,r^0)= z_a +
\eta E^u$ where $E^u$ is the unstable eigenvector, and $\eta$ is a
small parameter to be chosen below.  Let $t_c+1$ be the first time for
which the iterate of $z^0$ is beyond $\fix(R)$.  Now choose a point 
$z(\theta)$ on the line from $z^0$ to $z^1= f_\delta(z^{0})$, 
parameterized by the initial angle $\theta$.  By construction we are 
guaranteed that the function $ Z( \theta ) = \pi_{1} f^{t_c}_\delta 
(z(\theta))$ has a zero for $\theta \in [\theta^0, \theta^1]$.  We use 
a root finding method (Brent's method) to determine this zero to some 
precision, say $\rho$.  The choice of precision influences the 
original value for $\eta$, as well as the number of iterates until a 
crossing.  Assuming $W^{u}$ is smooth, the point $z^0$ will be 
$O(\eta^2)$ away from $W^u$.  After $t_c$ iterates, however, this 
error will decrease by the factor $\lambda^{-t_c}$ where $\lambda$ is 
roughly the unstable multiplier of the fixed point.  There is no sense 
in having this error smaller or larger than the precision of our root 
finder, so we set $\rho \sim \lambda^{-t_c} \eta^2$.  On the other 
hand, since we start a distance $\eta$ from the fixed point, and wish 
to go a distance $O(1)$ to find the first crossing of the symmetry 
line, we have $\eta \lambda^{t_c} = O(1)$.  Thus, it is appropriate to 
set $\eta \sim \rho^{1/3}$.  To find the second homoclinic point, $p 
\in \fix(f_\delta R)$, we repeat the above analysis, using crossing of 
$\fix(f_\delta R)$ to determine $t_c$, etc.  The lobe area is given by 
the difference in action between these two orbits, from \Eq{deltaW}.
	
For our computations, using IEEE double precision arithmetic, we set
$\rho = 10^{-19}$. These computations give apparently accurate results
providing $A >> 10^{-14}$. Subsequent to our obtaining these results, 
Delshams and Ram\'\i rez-Ros used extended precision arithmetic to 
obtain lobe areas as small as $10^{-4200}$ \cite{Delshams4}.

\Epsfig{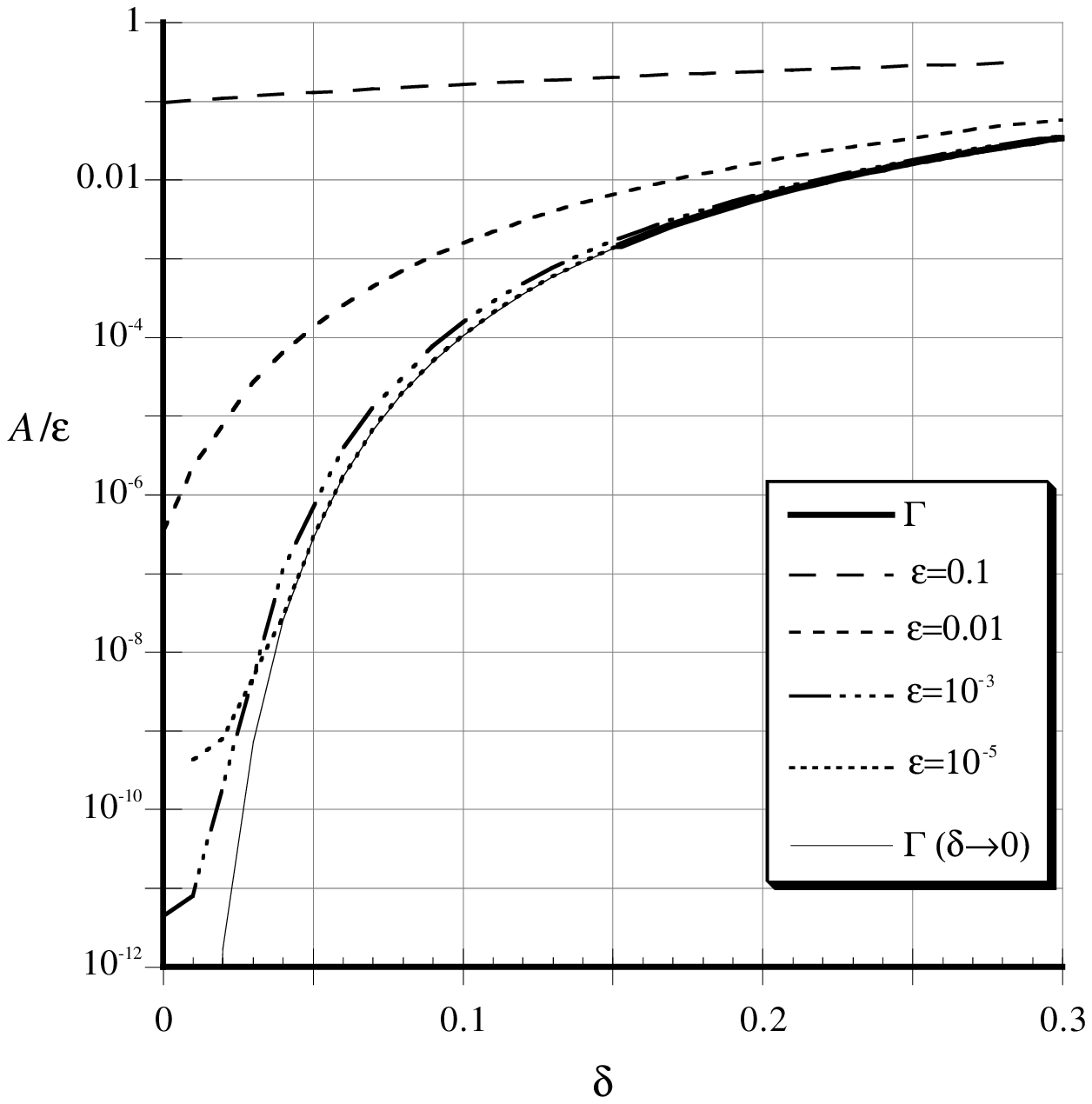}{ht}
       {Log-Linear plot of $A/\epsilon$ as a function of $\delta$ for 
		various values of $\epsilon$ compared with the theoretical 
		expression \Eq{theoarea} (thick line) and 
		the small $\delta$ expression \Eq{expsmall} (thin line)}
       {fig:logDW}{9 cm}
	   
In \Fig{fig:logDW} we show a comparison of the result of \Eq{theoarea}
with the numerical results on a log scale.  The analytical result
agrees well with the numerical results when $\epsilon = 10^{-5}$.  We
show the same data on a linear scale in \Fig{fig:AI-DW}.  We plot the
sum \Eq{gamma} only for $\delta > 0.15$, since it is indistinguishable
from asymptotic formula \Eq{expsmall} for smaller values.  Moreover,
the asymptotic formula agrees with the $\epsilon = 10^{-5}$
computation within $1\%$ up to $\delta=0.8$.  In \Fig{fig:AI-DW}, the
anti-integrable results evaluated at $\epsilon=1$ are also shown.  We
are unable to obtain numerical results for such a large $\epsilon$, as
the multiplier of the fixed points is too large.
\Epsfig{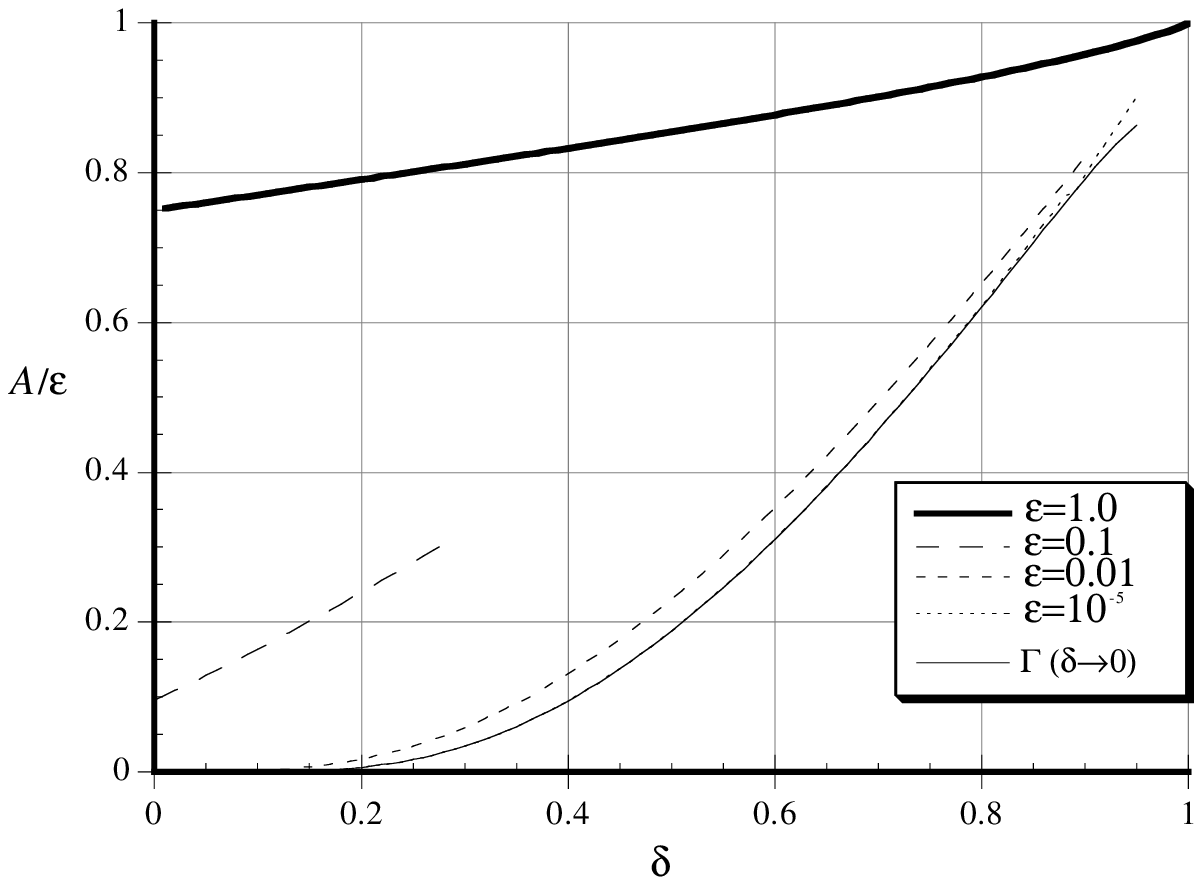}{ht}
       {Linear plot of $A/\epsilon$ as a function of $\delta$ for 
		various values of $\epsilon$. The solid curve is the anti-integrable 
		limit \Eq{antiI} for $\epsilon=1$.}
       {fig:AI-DW}{10 cm}
	   
\section{Conclusion}

The perturbed Suris map studied here depends on two parameters, the
Suris parameter $\delta$ and the perturbation strength $\epsilon$.  We
obtained the lobe area for the fixed point resonance of this map for
small $\epsilon$ to $O(\epsilon)$ and for large $\epsilon$ to
$O(\epsilon^{-1})$.  In the small epsilon case, the numerical
computations indicate that the asymptotic formula \Eq{expsmall} for
$\delta \to 0$ is a good approximation even up to $\delta = 0.8$.

A similar Melnikov analysis is possible for other standard maps 
\cite{Delshams1,Delshams3}.  In particular \cite{Lom2} showed that 
there exists a large class of standard maps with saddle connections.  
Interestingly these maps are not all integrable.  The use of the 
Melnikov potential is not restricted to twist maps; in fact, it can be 
applied to any exact symplectic map \cite{Delshams2}.

The method can also be applied to any higher dimensional twist map 
that has a saddle connection of the type described in this paper (see 
\cite{Delshams2, Lom0}).  Integrable examples of such maps have been 
found in \cite{McLac93,M-V,Sur2}, and the Melnikov method has been 
applied to applied to the four dimensional McLachlan map 
\cite{Delshams2}.  Similar techniques are also applicable to volume 
preserving maps \cite{Lom3}.  The study of perturbations of twist 
maps with saddle connections in higher dimensions is important because 
it could help in the development of a higher dimensional theory of 
transport.

\bibliographystyle{unsrt}
\bibliography{suris}
\end{document}